\documentclass[10pt,final]{article}

\usepackage[english]{babel}[1999/09/09]

\usepackage[english]{varioref}

\usepackage{tbook}

\usepackage{tbook-pl}
\ifgallery\else
\ifpdflatex\usepackage[%
  pdftex,pagebackref=true,pdfstartview=FitH,
  pdfcreator={tbook XML + LaTeX with hyperref package},
  pdfpagelayout=OneColumn,
  pdfpagelabels=true,
  draft=false,plainpages=false]{hyperref}[2000/05/08]
  \hypersetup{%
    pdftitle={ReacProc: A Tool to Process Reactions Describing Particle Interactions},
    pdfauthor={Andrey SiverIHEP},
    pdfkeywords={particles properties, conservation laws, Particle Data Group, PPDL, ReacProc(v0.0.9)}
  }
\IfFileExists{hypcap.sty}{\usepackage{hypcap}}{}
\IfFileExists{pdfcolmk.sty}{\usepackage{pdfcolmk}}{}
\fi\fi

\iftbookcfg\input tbook.cfg\fi
\begin{document}
   \title{ReacProc: A Tool to Process Reactions Describing Particle Interactions}
 \author{Andrey~Siver\thanks{IHEP}}

\maketitle
\tableofcontents

\begin{abstract}

\par ReacProc is a program written in C/C++ programming language which 
can be used (1) to check out of reactions describing particles interactions 
against conservation laws and (2) to reduce input reaction into some canonical form. Table with particles properties is available within ReacProc package.

\end{abstract}

\section{Introduction}

\par ReacProc can be used (1) to check out of reactions describing particles interactions against conservation laws (for example, for electric charge) and (2) to reduce input reaction into some canonical form.
\smallskip \par  The followings functionalities were realized:

\begin{enumerate}

\item \label{parse}%
Parsing of reaction. It means reading reaction from a stream and building corresponding dynamical structure of connection lists. Also reaction syntax is validating against some grammar (see section \REF{}{grama});
\item \label{reduce}%
Reducing of grouping brackets (see section \REF{}{desc});
\item \label{synony}%
Particle synonyms processing (see section \REF{}{desc});
\item \label{sort}%
Sorting of particle and particle groups in reaction (according to an order in a dictionary) (see section \REF{}{desc});
\item Checking conservation laws and printing messages if it'{}s needed (see section \REF{}{desc});
\item \label{output}%
Print reaction to output stream from corresponding data structure (see section \REF{}{desc}).\end{enumerate}

\par We use PPDL notation [\citealp{primer},\citet{ppdl}] for reaction.
\par Reducing to a canonical form means applying \REF{}{parse}, \REF{}{reduce}, \REF{}{synony}, \REF{}{sort}, \REF{}{output} functionalities correspondingly.

\par ReacProc is written in C/C++ programming language and uses  Bison[\citealp{bison}]+Flex[\citealp{flex}] parser generator.
\par In order to check conservation laws we need some table with particle properties. This file was created from publicly available Particle Data Group file and contains following particle properties: electric charge, barion number, S-{}, C-{}, B-{}, T-{}quantum numbers. There was also created file with lepton quantum numbers. ReacProc can check conservation laws corresponding to all of these properties. There is also file containing a dictionary.
\par  To run ReacProc one do call corresponding executable with input file. ReacProc reads line by line and process strings ending with {\mbox{\char34{}}};{\mbox{\char34{}}}. {\mbox{\char34{}}}{\mbox{\char92{}}}{\mbox{\char34{}}} symbol is considered as line continuation character (multi-{}lines can be written with it). During processing ReacProc creates several files:

\begin{itemize}

\item {\mbox{\char34{}}}rp-{}accept.txt{\mbox{\char34{}}} contains reactions satisfying tested conservation laws (in {\mbox{\char34{}}}rp-{}accept-{}s.txt{\mbox{\char34{}}} reactions are sorted);
\item {\mbox{\char34{}}}rp-{}reject.txt{\mbox{\char34{}}} contains reactions which do not satisfy tested conservation laws (in {\mbox{\char34{}}}rp-{}reject-{}s.txt{\mbox{\char34{}}} reactions are sorted);
\item {\mbox{\char34{}}}rp-{}unknown.txt{\mbox{\char34{}}} contains reactions which cannot be processed by ReacProc;
\item {\mbox{\char34{}}}rp-{}log.txt{\mbox{\char34{}}} contains debug messages and additional information about processed reactions.\end{itemize}

\par Another kind of usage ReacProc (and perhaps more powerful) is usage it as a module in another system (for example, in specialized databases). So it'{}s important to describe general structure of ReacProc. The rest of the paper is devoted to this point.
\par As an example of application let'{}s call possible program for building mathematical observables from corresponding reactions (which in their order can be used in some calculations).
\par ReacProc available here \url{http://sirius.ihep.su/~siver/rp.zip}

\section{Data structures}

\par The most important data structure is called '{}f{\mbox{\char95{}}}s'{}. Here is its definition written in C programming language:

\medskip \par \begin{varverbatim}
struct f{\mbox{\char95{}}}s \{ 			
	struct f{\mbox{\char95{}}}s *l;
	struct f{\mbox{\char95{}}}s *r;
	int c;
	char *n;
	struct gf{\mbox{\char95{}}}s *deca;
\}
\end{varverbatim}
, where the fields means: '{}l'{},'{}r'{} -{} pointers to abstract left and right (or previous and next) '{}f{\mbox{\char95{}}}s'{} elements, '{}n'{} -{} pointer to name of particle, '{}c'{} -{} counter of particles, '{}deca'{} -{} pointer to decay description of particle. So, this structure is for one type of particle appearing in reaction. But since it can be connected with others such structure, any reference to this structure will be reference to the group of particles structures.
\smallskip \par  '{}gf{\mbox{\char95{}}}s'{} data type is defined this way:

\medskip \par \begin{varverbatim}
struct gf{\mbox{\char95{}}}s \{			
	struct gf{\mbox{\char95{}}}s *l;
	struct gf{\mbox{\char95{}}}s *r;
	struct f{\mbox{\char95{}}}s *pg;
	int sign;
\}
\end{varverbatim}
, where the fields means: '{}l'{},'{}r'{} -{} pointers to abstract left and right (or previous and next) '{}gf{\mbox{\char95{}}}s'{} elements, '{}pg'{} -{} reference to particle group, '{}sign'{} -{} unary sign which prescribing to '{}pg'{} group.

\medskip \par Full reaction is represented as following:

\medskip \par \begin{varverbatim}
struct reac \{
	struct f{\mbox{\char95{}}}s *is;
	struct gf{\mbox{\char95{}}}s *fs;
\}
\end{varverbatim}
, where '{}is'{} field stands for initial state (sequence of particles types, ordinary in amount of two or one), '{}fs'{} field stands for final state (sequence of groups of particle sequences).

\section{Description of the basic functions\label{desc}%
}

\par Let'{}s describe most important functions and global variables which they use.
\begin{itemize}

\item 
\par The parsing function:

\par \begin{varverbatim}int yyparse(void);\end{varverbatim}
	Description: this function is generated by Bison. It calls '{}yylex'{} function which in its order reads reaction from '{}yyin'{} stream. After call '{}yyparse'{} reaction will be in '{}r'{} global variable.

\item 
\par The canonization function:

\par \begin{varverbatim}struct reac* canoni(reac* r);\end{varverbatim}
	Description: it transforms reaction '{}r'{} into the canonical form. It calls '{}syn{\mbox{\char95{}}}proc{\mbox{\char95{}}}gf{\mbox{\char95{}}}s'{}, '{}reduce{\mbox{\char95{}}}gf{\mbox{\char95{}}}s'{} and '{}sort{\mbox{\char95{}}}full{\mbox{\char95{}}}gf{\mbox{\char95{}}}s'{} functions.

\item \label{syn_proc}%

\par The particle synonims processing function:
\begin{varverbatim}f{\mbox{\char95{}}}s* syn{\mbox{\char95{}}}proc{\mbox{\char95{}}}gf{\mbox{\char95{}}}s(f{\mbox{\char95{}}}s*);\end{varverbatim}
	Description: it replaces synonim of particle by the key synonim (which actually is PDG name) which stands at first place in synonims list of the dictionary file;

\item \label{reduce_gf_s}%

\par The function for round brackets (named {\mbox{\char34{}}}grouping{\mbox{\char34{}}}) reducion:
\begin{varverbatim}gf{\mbox{\char95{}}}s* reduce{\mbox{\char95{}}}gf{\mbox{\char95{}}}s(gf{\mbox{\char95{}}}s*);\end{varverbatim}
	Description: it reduces parenthesis. For example, expression '{}( E+ + MU+ ) X'{} is equal to '{}E+ X + MU+ X'{}. Expression with grouping brackets is represented as {\mbox{\char34{}}}generalized{\mbox{\char34{}}} particle with '{}n'{} and '{}c'{} fields equaling to {\mbox{\char34{}}}?{}compo{\mbox{\char34{}}} and 0 correspondingly;  

\item \label{sort_full_gf_s}%

\par The sorting function:
\begin{varverbatim}gf{\mbox{\char95{}}}s* sort{\mbox{\char95{}}}full{\mbox{\char95{}}}gf{\mbox{\char95{}}}s(gf{\mbox{\char95{}}}s*);\end{varverbatim}
	Description: it sorts according to '{}ord{\mbox{\char95{}}}full{\mbox{\char95{}}}gf{\mbox{\char95{}}}s'{} and at the lowest level -{} '{}ord(char*,char*)'{} function which has two ordering modes: lexicographical (in case a global '{}ord{\mbox{\char95{}}}type'{} = '{}ORD{\mbox{\char95{}}}TRUE{\mbox{\char95{}}}LEX'{}) that is order that every English dictionary has, and so called dictionary order (in case the global '{}ord{\mbox{\char95{}}}type'{} = '{}ORD{\mbox{\char95{}}}DICT'{}) that is order of some particle dictionary (ReacProc uses '{}dict-{}syn.txt'{} file as a particle dictionary by default). These modes have advantages in certain situations: while ORD{\mbox{\char95{}}}TRUE{\mbox{\char95{}}}LEX mode is useful when we compare reactions, with ORD{\mbox{\char95{}}}DICT mode sorting is more logical and physical well founded (see example below);

\item \label{test_reac}%

\par The function checking conservation law:
\begin{varverbatim}int test{\mbox{\char95{}}}reac(reac *r, const char s[], prop *pp, int kolvo);\end{varverbatim}
	Description: it tests reaction '{}r'{} against law naming '{}s'{} and  property table '{}pp'{} with '{}kolvo'{} elements. It return '{}0'{} if no error was detected and some signal (such as RP{\mbox{\char95{}}}SIGN{\mbox{\char95{}}}P{\mbox{\char95{}}}UNK) otherwise. It uses global variable '{}SIGN{\mbox{\char95{}}}conser'{} which assigns the type of conservation law (0 for equality, -{}1 and 1 for '{}<{}='{} and '{}>{}='{} equality respectively.

\item 
\par The output function: 
\begin{varverbatim}void out{\mbox{\char95{}}}to(FILE *f, reac* s);\end{varverbatim}
	Description: it prints reaction '{}s'{} into stream '{}f'{}. It uses '{}out{\mbox{\char95{}}}f{\mbox{\char95{}}}s'{} and '{}out{\mbox{\char95{}}}gf{\mbox{\char95{}}}s'{} functions.
\end{itemize}

\par Let'{}s write a reaction and see how it'{}s represented.
\begin{varverbatim}
	E+ E-{} -{}-{}>{} W-{} <{} QUARK QUARKBAR >{} W+ <{} TAU+ NUTAU + MU+ NUMU + E+ NUE >{}\end{varverbatim}
	, where brockets denote sub-{}decays processes.
	Using the labels (Ax, Bx, Cx, Dx, Ex) instead of memory pointers memory structure corresponding to the reaction can be shown by following ('{}-{}>{}'{} sign shows value of the variables and '{}-{}-{}>{}'{} shows value of the references)
\begin{varverbatim}
	L1: reaction ( is -{}>{} I1 , fs -{}>{} F1 );
	
	I1: f{\mbox{\char95{}}}s ( l -{}>{} NULL , r -{}>{} I2 , c -{}>{} 1 , n -{}-{}>{} {\mbox{\char34{}}}E+{\mbox{\char34{}}} , deca -{}>{} NULL);	
	I2: f{\mbox{\char95{}}}s ( l -{}>{} I1 , r -{}>{} NULL , c -{}>{} 1 , n -{}-{}>{} {\mbox{\char34{}}}E-{}{\mbox{\char34{}}} , deca -{}>{} NULL);
	F1: gf{\mbox{\char95{}}}s ( l -{}>{} NULL , r -{}>{} NULL , pg -{}>{} F2 , sign -{}>{} +1 );
	F2: f{\mbox{\char95{}}}s ( l -{}>{} NULL , r -{}>{} F3 , c -{}>{} 1 , n -{}-{}>{} {\mbox{\char34{}}}W-{}{\mbox{\char34{}}} , deca -{}>{} A1);	
	F3: f{\mbox{\char95{}}}s ( l -{}>{} F2 , r -{}>{} NULL , c -{}>{} 1 , n -{}-{}>{} {\mbox{\char34{}}}W+{\mbox{\char34{}}} , deca -{}>{} B1);	
	A1: gf{\mbox{\char95{}}}s ( l -{}>{} NULL , r -{}>{} NULL , pg -{}>{} A2 , sign -{}>{} +1 );
	A2: f{\mbox{\char95{}}}s ( l -{}>{} NULL , r -{}>{} A3 , c -{}>{} 1 , n -{}-{}>{} {\mbox{\char34{}}}QUARK{\mbox{\char34{}}} , deca -{}>{} NULL);
	A3: f{\mbox{\char95{}}}s ( l -{}>{} A2 , r -{}>{} NULL , c -{}>{} 1 , n -{}-{}>{} {\mbox{\char34{}}}QUARKBAR{\mbox{\char34{}}} , deca -{}>{} NULL);
	B1: gf{\mbox{\char95{}}}s ( l -{}>{} NULL , r -{}>{} B2 , pg -{}>{} C1 , sign -{}>{} +1 );			
	B2: gf{\mbox{\char95{}}}s ( l -{}>{} NULL, r -{}>{} B3 , pg -{}>{} D1 , sign -{}>{} +1 );
	B3: gf{\mbox{\char95{}}}s ( l -{}>{} B2 , r -{}>{} NULL , pg -{}>{} E1 , sign -{}>{} +1 );	
	C1: f{\mbox{\char95{}}}s ( l -{}>{} NULL , r -{}>{} C2 , c -{}>{} 1 , n -{}-{}>{} {\mbox{\char34{}}}TAU+{\mbox{\char34{}}} , deca -{}>{} NULL);
	C2: f{\mbox{\char95{}}}s ( l -{}>{} C1 , r -{}>{} NULL , c -{}>{} 1 , n -{}-{}>{} {\mbox{\char34{}}}NUTAU{\mbox{\char34{}}} , deca -{}>{} NULL);
	D1: f{\mbox{\char95{}}}s ( l -{}>{} NULL , r -{}>{} D2 , c -{}>{} 1 , n -{}-{}>{} {\mbox{\char34{}}}MU+{\mbox{\char34{}}} , deca -{}>{} NULL);
	D1: f{\mbox{\char95{}}}s ( l -{}>{} D1 , r -{}>{} NULL , c -{}>{} 1 , n -{}-{}>{} {\mbox{\char34{}}}MUTAU{\mbox{\char34{}}} , deca -{}>{} NULL);
	E1: f{\mbox{\char95{}}}s ( l -{}>{} NULL , r -{}>{} E2 , c -{}>{} 1 , n -{}-{}>{} {\mbox{\char34{}}}E+{\mbox{\char34{}}} , deca -{}>{} NULL);
	E2: f{\mbox{\char95{}}}s ( l -{}>{} E1 , r -{}>{} NULL , c -{}>{} 1 , n -{}-{}>{} {\mbox{\char34{}}}NUE{\mbox{\char34{}}} , deca -{}>{} NULL);		
	\end{varverbatim}
	After canonization in the case of '{}ord{\mbox{\char95{}}}type=ORD{\mbox{\char95{}}}TRUE{\mbox{\char95{}}}LEX'{} the reaction has following form :
\begin{varverbatim}e+ e-{} -{}-{}>{} W+ <{} e+ nu(e) + mu+ nu(mu) + nu(tau) tau+  >{} W-{} <{} QUARK QUARKBAR >{} ;\end{varverbatim}
	and in the case of '{}ord{\mbox{\char95{}}}type=ORD{\mbox{\char95{}}}DICT'{}:
\begin{varverbatim}e+ e-{} -{}-{}>{} W+ <{} e+ nu(e) + mu+ nu(mu) + tau+ nu(tau)  >{} W-{} <{} QUARK QUARKBAR >{} ;\end{varverbatim} (notice order of '{}tau+'{} and '{}nu(tau)'{}).

\section{ Appendix A. The grammar of reactions\label{grama}%
}

\par This appendix contains reactions grammar using in ReacProc. Entry point is '{}reaction'{}. It was cut out from corresponding Bison *.y file.
\bigskip \par 
\begin{varverbatim}
reaction	:	initial{\mbox{\char95{}}}state ARROW final{\mbox{\char95{}}}state END
		|      	error END
		;
initial{\mbox{\char95{}}}state	:	beam target
		;
beam		:	PARTICLE
		;
target		:	/* empty */
		|	PARTICLE target
		;
particle	:	PARTICLE
		|	PARTICLE LC final state RC
		|	LP final state RP
		;
decay{\mbox{\char95{}}}group	:	/* Empty */			// d.{\mbox{\char95{}}}g. has type f{\mbox{\char95{}}}s
		|	particle decay{\mbox{\char95{}}}group
		;
final{\mbox{\char95{}}}state	:	decay{\mbox{\char95{}}}group			// f.s. has type gf{\mbox{\char95{}}}s
		|	decay{\mbox{\char95{}}}group PLUS final state
		|	decay{\mbox{\char95{}}}group MINUS final state
		|	final state PLUS CC
\end{varverbatim}

\bigskip \par Terminal elements (such as ARROW, END and etc) are described below. This Flex code was cut out from corresponding *.lex file:
\bigskip \par 
\begin{varverbatim}
{\mbox{\char34{}}}-{}-{}>{}{\mbox{\char34{}}}				return(ARROW);
{\mbox{\char34{}}};{\mbox{\char34{}}}					return(END);
{\mbox{\char34{}}}({\mbox{\char34{}}}					return(LP);
{\mbox{\char34{}}}){\mbox{\char34{}}}					return(RP);
{\mbox{\char34{}}}<{}{\mbox{\char34{}}}				return(LC);
{\mbox{\char34{}}}>{}{\mbox{\char34{}}}				return(RC);
{\mbox{\char34{}}}+{\mbox{\char34{}}}					return(PLUS);
{\mbox{\char34{}}}-{}{\mbox{\char34{}}}					return(MINUS);
{\mbox{\char34{}}}CC{\mbox{\char34{}}}					return(CC);
[ {\mbox{\char92{}}}t{\mbox{\char92{}}}n{\mbox{\char92{}}}r]+			;		/*	Do nothing	*/            
[A-{}Za-{}z{\mbox{\char92{}}}+{\mbox{\char92{}}}-{}{\mbox{\char92{}}}:{\mbox{\char92{}}}{\mbox{\char34{}}}{\mbox{\char92{}}}.{\mbox{\char92{}}}*{\mbox{\char92{}}}={\mbox{\char92{}}}\%{\mbox{\char92{}}}{\mbox{\char95{}}}{\mbox{\char92{}}}({\mbox{\char92{}}})0-{}9{\mbox{\char92{}}}/]+ 	return(PARTICLE);	
\end{varverbatim}

\section{Acknowledgments}

\par Author would like to thank Zenin O. V., Ezhela V. V.
\par The work was \textbf{not} supported by the project RFFI-{}05-{}07-{}90191-{}w.\long\def\bibpreamble{\parindent10pt%

}
\bibliographystyle{tbenl}
 \citeindexfalse   
 \bibliography{biblio}

\end{document}